\documentclass[aps,prl,preprint,groupedaddress]{revtex4}

\usepackage{graphicx}

\begin{document}


\title{Scattering of Ultra Cold Neutrons on  Nano-size Bubbles}


\author{Vladimir Gudkov }
\email{ gudkov@sc.edu}
\affiliation{Department of Physics and Astronomy
\\ University of South Carolina \\
Columbia, SC 29208 }


\date{\today}

\begin{abstract}
Inelastic scattering of ultra cold neutrons on bubbles with the size of nanometers is considered. It is shown that neutron-bubble cross section is large and sensitive to different vibration modes of bubbles. This process could be used for study of dynamics of nano-size bubbles, and for new methods of ultra cold neutron production if appropriately sized bubbles with sufficient density can be created.
\end{abstract}


\maketitle

\section{Introduction}

The understanding of interactions of ultra cold neutrons (UCN) with materials is very important for many applications (see, for example \cite{golub} and references therein) including fundamental neutron physics, neutron scattering, and condense matter physics . A special interest in interactions of UCN with nano size particles as an explanation of the possible loss of UCNs during a storage in a trap and as a possible new method for UCN production has been emphasized in papers \cite{vn1,vn2}. A calculation of neutron interactions with large particles is a very complicated problem since it requires detailed description of internal excitation modes for these particles, which is impossible in general, and can be done only numerically for some cases. To avoid these difficulties in general understanding of properties of interaction of UCN with nano-particles, we consider a bubble of nano-size in liquid for which the neutron scattering problem may be solved analytically. In particular, we consider bubbles in liquid helium which is a ``natural'' environment for UCN applications.

Let us recall that a characteristic  scale of kinetic energy of UCN could be as low as at the level of nano-electronvolts. These energies corresponds to neutron wavelengths at the nano-meter scale. For example, neutron with the kinetic energy  $E_n=100\; neV$ has a wavelength  $\lambda = 90.4\;  nm$, and the energy $E_n=1\; neV$ corresponds to the neutron wavelength $\lambda = 904\;  nm$. Then a typical neutron scattering parameter on the nano-size particles is  $(kR)\sim 0.07 - 10$, where $k=\sqrt{2mE_n}/\hbar$ is neutron wave vector and $R$ is size of a target. This parameter gives a   characteristic range of partial waves required for the cross sections calculations.
Therefore, in general case UCN cross sections with nano-particles can not be calculated in the low energy approximations ($kR\ll 1$), but contributions from many partial waves  should be taken into account. In other words, very slow neutrons can demonstrate ``high energy'' scattering properties when they interact with nano-particles.

In this letter, we present results of calculations of inelastic neutron cross sections on the bubble of the nano-meter sizes. Some useful results for elastic cross sections are discussed briefly since all details for elastic scattering could be found in quantum mechanics textbooks.

\section{Elastic scattering on spherical targets}
For calculation of elastic neutron scattering on the large objects we can use similarity between cold neutron scattering and light scattering on transparent objects in optics. In our case, it is convenient to use neutron Fermi potential which is related to refractive index in optics. The values of Fermi potential for a medium with an atomic density $N$ and neutron coherent scattering length $b_0$ is\cite{gurev,golub}
\begin{equation}\label{fermi}
    V=\frac{2\pi \hbar^2}{m}N b_0,
\end{equation}
where $m$ is a neutron mass. This gives Fermi potential for neutron propagation in liquid $^4He$, $^3He$ and solid deuterium as $16\; neV$, $28\; neV$ and $104\;neV$ \cite{liqdeut}, correspondingly. Therefore, ultra cold neutrons in helium can feel a potential steps or a potential well of  about  $\sim 20\, neV$ ($\sim 80\, neV$), when they are scattering on bubbles (deuterium spheres), correspondingly. One can see that for neutron elastic scattering in this approach there is no differences between neutron scattering on bubbles and on spherical solid objects, which is not the case for inelastic scattering.  To describe this scattering process, one can use a standard results for scattering on square-well potential $V_0$. Then two separate cases with essentially different scattering regimes can be considered: a ``high energy'' regime where parameter $kR\sim 1$ and a ``low energy'' regime where $kR\ll 1$.
The first case results in a total cross section as a sum over all important partial waves with phase shifts $\delta_l$ for angular momenta $l\leq L_{max}$:
\begin{equation}\label{sumpw}
    \sigma = \frac{4\pi}{k^2}\sum^{L_{max}}_{l=0} (2l+1)\sin ^2 \delta_l,
\end{equation}
where
\begin{equation}\label{tandel}
    \tan \delta_l = \frac{j^{\prime}_l(kR)- D j_l(kR) }{n^{\prime}_l(kR)- D n_l(kR) },
\end{equation}
and
\begin{equation}\label{tandel}
D = \frac{k_0}{k}\frac{j^{\prime}_l(k_0R)}{j_l(k_0R) }.
\end{equation}
Here $j_l(z)=\sqrt{\pi/(2z)}J_{l+1/2}(z)$ and $n_l(z)=\sqrt{\pi/(2z)}N_{l+1/2}(z)$ are spherical Bessel functions, and $k_0=\sqrt{2m(E_n+V_0)}/\hbar$.

\begin{figure}
 \includegraphics{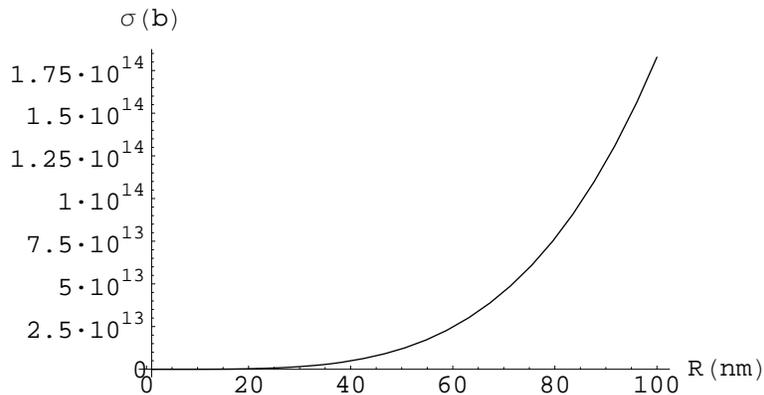}%
 \caption{Elastic total cross section of neutrons with energy $E_n=100\; neV$ as a function of a bubble radius in liquid helium. }
 \label{e100}
 \end{figure}

\begin{figure}
 \includegraphics{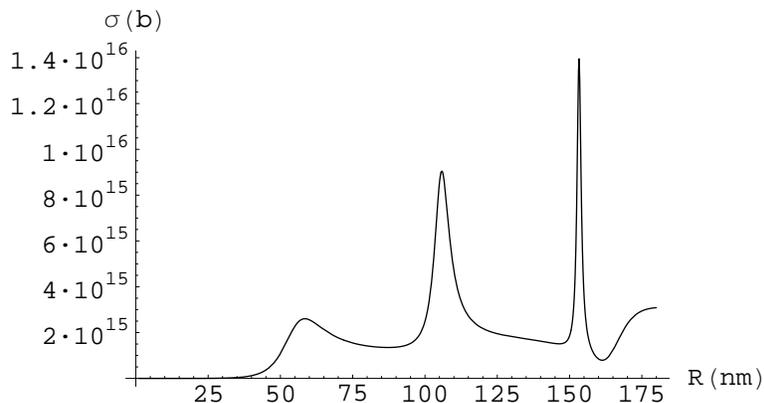}%
 \caption{Elastic total cross section of neutrons with energy $E_n=1\; neV$ as a function of a bubble radius in liquid helium.}
\label{e1}
 \end{figure}

In spite of the fact that the potential well is extremely shallow, the size of the well can be rather large. Therefore, neutron scattering can show resonance behavior due to a sensitivity to a possible virtual (or real) bound states (for  neutron bound states in matter see \cite{bound1,bound2,bound3}) in large bubbles when $E_n\leq V_0$ and $2mV_0R^2/\hbar^2 \sim 1$. Numerical calculations confirm such phenomena:  the cross section as a function of the size of the target   is a smoothly growing function when neutron energy  is much larger than the potential (for example, for $V_0=16 \; neV$ and $E_n=100 \; neV$, see fig. (\ref{e100})), but it shows irregular (resonance) behavior when neutron energy  is much smaller  than the potential (for example, $E_n=1 \; neV$,  see fig. (\ref{e1})). (For the considered parameters $V_0$,  $E_n$ and $\lambda \sim R$ it was required to take into account contributions from high orbital momenta up to $L_{max} \sim 20$.)

For a ``low energy'' regime with $(kR)\ll 1$, one can consider only zero orbital momentum contribution using low energy expansion for the scattering phase. Then, the cross section is
\begin{equation}\label{elcs}
    \sigma \sim \frac{4}{9}R^2(K_0R)^4 \sim R^6 \sim N_t^2,
\end{equation}
where $K_0=\sqrt{2mV_0}/\hbar$, and $N_t$ is a total number of nuclei in the target (or the total number of nuclei which could  fill up the volume of the bubble). One can see a coherent enchantment factor $N_t$ for elastic scattering cross section. However, the condition required to obtain Eq.(\ref{elcs}), $(kR)\ll 1$ is equivalent to the condition $\lambda \gg R$. Therefore, it does not lead to a large value of  coherent  cross section in compare to a geometrical one (it is  suppressed by the factor $(K_0R)^4$ as seen from Eq.(\ref{elcs})).

Since that elastic cross section for scattering of UCN on nano-particles shows regular and expected properties, one can conclude that it is  not large enough to be considered for the mechanisms of UCN energy variations suggested in \cite{vn1,vn2}) where ``a sufficiently large cross section for coherent interaction'' \cite{vn1} was required.

\section{Inelastic scattering on bubbles}

Inelastic scattering of UCN on a bubble in a liquid  can be calculated using a general formula for inelastic cross sections \cite{landau3} with a transition of the target from an initial $i$ to a final $f$ state:
\begin{equation}\label{land}
   d\sigma_{fi} = \frac{m p^{\prime}}{4 \pi ^2 \hbar ^4}|\langle f p^\prime |U|i p \rangle|^2 d\Omega ^{\prime}.
\end{equation}

Assuming that energy of  neutrons is high, compared with the neutron binding energy in the bubble, one can approximate incoming and outgoing neutron wave functions by plane waves.  It should be noted that this approximation is always \cite{landau3} correct if $|U|\ll \hbar^2/ m R^2$, where $R$ is a radius the bubble. Therefore, it can be safely used for bubbles with radii less than tens of nanometers irrespectively to the binding energy condition. Then, the cross section can be written as
\begin{equation}\label{cs}
   d\sigma_{fi} = \frac{m^2 p^{\prime}}{4 \pi ^2 \hbar ^4 p}\left|\int \int U e^{-i \vec{q} \vec{r}}\psi^*_f\psi_i d\tau d\vec{r}\right|^2 d\Omega ^{\prime},
\end{equation}
where $\psi_f (\tau )$ and $\psi_i (\tau )$ are final and initial wave functions of the bubble, and $\vec{q} = \vec{k^\prime} - \vec{k}$ is transferred wave vector ($\vec{k^\prime}=\vec{p^\prime}/\hbar$ and $\vec{k}=\vec{p}/\hbar$). It is convenient to re-write the cross section in terms of $q$ using the identity $qdq=(kk^\prime /2 \pi)d\Omega ^{\prime}$. Then, taking into account that the potential inside the bubble has a constant value $V_0$, one obtains
\begin{equation}\label{csf}
   d\sigma_{fi} = \frac{m^2 V_0^2 }{2 \pi  \hbar ^4 k^2}\left|\int \int e^{-i \vec{q} \vec{r}}\psi^*_f\psi_i d\tau d\vec{r}\right|^2 q dq,
\end{equation}
where integration over $\vec{r}$ is performed inside the volume of the (deformed) bubble. (Since we consider neutron propagation in medium, it is natural to assume the reference zero Fermi potential for the medium. Therefore the only   non-zero effective potential is the potential inside the bubble volume.) The shape of the bubble can be described  in spherical coordinates by equation \cite{bm}
\begin{equation}\label{shape}
    R=R_0+\sum_{l m}a_{lm} Y_{lm}^*(\theta , \phi),
\end{equation}
where $R_0$ is a radius of a spherical (non-deformed) bubble, $a_{lm}$ are deformation parameters in an expansion by spherical functions $Y_{lm}(\theta , \phi)$. Assuming small deformations ($a_{lm} \ll R_0$), one can describe dynamics of the bubble in terms of harmonic oscillators in the  space of  deformation parameters \cite{bm}.
Then, the integrant for the integration over the space of deformation $\tau$ in the Eq.(\ref{csf}) is proportional to the product the first order of the corresponding deformation parameter, and harmonic oscillator functions $\psi_{i=n}$ and $\psi_{f=n\pm 1}$, which results in $\sqrt{\hbar n/2 M \omega}$ for the integral over $\tau$. Here $n$ is an integer number, M is a effective mass of the bubble, and $\omega$ is a frequency of the bubble oscillation.

In this paper, we restrict ourselves by first simple modes of the possible deformations: the breathing mode ($l=0$) and the surface mode ($l=2$). (The dipole mode with $l=1$ describes a shift of the bubble in the space and does not lead to a deformation.) For these two modes, one can calculate integrals over the $\vec{r}$ space ($ \int e^{-i \vec{q} \vec{r}}d\vec{r}$)
inside of the deformed bubble with the bubble shape described by Eq.(\ref{shape}). This integral is coupled with harmonic oscillator integral by deformation parameters. The harmonic oscillator integral is not equal to zero if it is contains linear terms  of deformation parameters. Therefore, one can expand the $e^{-i \vec{q} \vec{r}}$ function and keep only the part which is proportional to the first order of deformation (linear term) in the expansion.   Then the matrix element in Eq.(\ref{csf}) can be written as
\begin{equation}\label{matr}
 {\sl M} = \int \int e^{-i \vec{q} \vec{r}}\psi^*_f\psi_i d\tau d\vec{r}=R_0^2\sqrt{\frac{\hbar n}{2 M \omega}}\Phi_{mode} (\xi),
\end{equation}
where $\xi = q R_0$.  The function $\Phi_{b} (\xi)$ for the breathing mode is
\begin{equation}\label{fbr}
    \Phi_{b} (\xi) = 4 \pi \frac{\sin \xi}{\xi },
\end{equation}
and the function $\Phi_{s} (\xi)$ for the surface mode is
\begin{equation}\label{fsf}
    \Phi_{s} (\xi) = \frac{2\sqrt{5\pi}}{\xi^3 }[3 \xi\cos \xi -(3-\xi^2)\sin \xi ].
\end{equation}

 The Eq.(\ref{fbr}) is a direct result of integration over bubble volume modified by the first term in the radius expansion of Eq.(\ref{shape}). To obtain Eq.(\ref{fsf}), a simple symmetric ellipsoidal shape ($a_{22}=a_{2-2}$ and $a_{21}=a_{2-1}=0$) of the bubble deformation was assumed. However, the assumption about symmetry of deformation is not important for our considerations because different shapes may lead  only to slightly different coefficients in functions $\Phi$.

Then, differential cross sections for inelastic scattering of UCN on  bubbles can be written as
\begin{equation}\label{dif}
   d\sigma_{n\leftarrow n-1}(\xi ) = \frac{m^2 V_0^2 R_0^2}{4 \pi  \hbar ^4 k^2}\left( \frac{\hbar n}{M\omega }\right)\Phi_{mode}^2 (\xi)\xi d\xi
\end{equation}
for the case when energy transfers from the neutron to the bubble (down-scattering cross section), and as
\begin{equation}\label{difup}
   d\sigma_{n\leftarrow n+1}(\xi ) = \frac{m^2 V_0^2 R_0^2}{4 \pi  \hbar ^4 k^2}\left( \frac{\hbar (n+1)}{M\omega }\right)\Phi_{mode}^2 (\xi)\xi d\xi
\end{equation}
for the case when energy transfers from the bubble to the neutron (up-scattering cross section). It should be noted that due to rotational invariance these cross sections are independent on $m$-number in Eq.(\ref{shape}). For example, for the case of Eq.(\ref{fsf}) this independence comes automatically  after integration over angular parameter $\phi$.
The total cross section could be  obtained by integration of $d\sigma(\xi )$ from $\xi_{min} $ until $\xi_{max}$, taking into account that $ k^2 - {k^{\prime}}^2 = \pm 2 m \omega / \hbar$, where ``+'' at the right side corresponds to the case of Eq.(\ref{dif}), and ``-''  for  Eq.(\ref{difup}).

\begin{figure}
 \includegraphics{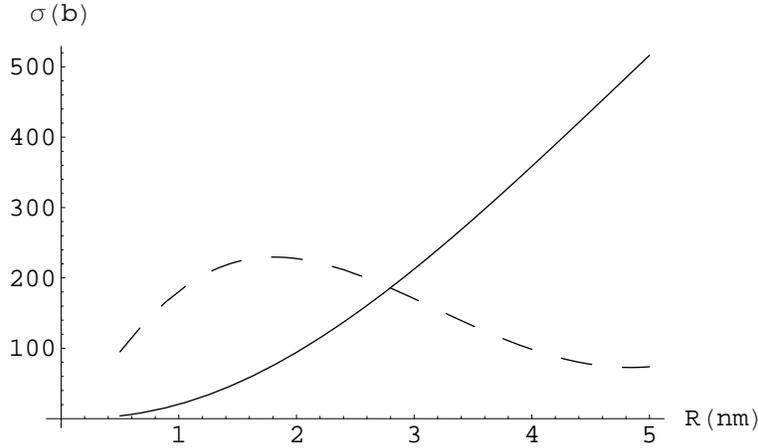}%
 \caption{Inelastic total up-scattering cross sections of neutrons with energy $E_n=100; neV$ as a function of a bubble radius in liquid helium for breathing (dashed line) and surface (solid line)  modes. ($n=1$)
 \label{ener100}}
 \end{figure}

\begin{figure}
 \includegraphics{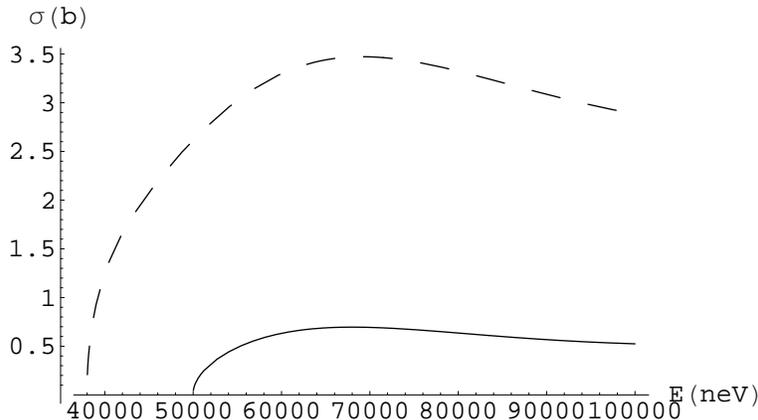}%
 \caption{Inelastic total down-scattering cross sections on the bubble of radius of $1.75; nm$ in liquid helium as a function of neutron energy for  breathing (dashed line) and surface (solid line)  modes. ($n=1$)\label{r175}}
 \end{figure}

Using results of papers \cite{bl1,bl2,bl3,bl4,bl5}, the frequency of the breathing mode can be written as
\begin{equation}\label{frqb}
    \omega_b ^2= \beta /2 \pi \rho R_0^3,
\end{equation}

the frequency of the quadrupole surface modes mode as
\begin{equation}\label{frqs}
    \omega_s ^2= 12 \alpha /\rho R_0^3,
\end{equation}
and the effective mass of the bubble as
\begin{equation}\label{frqs}
  M= 4 \pi \rho R_0^3.
\end{equation}
Here $\beta$ is a constant dependant on the properties of the liquid, $\alpha$ is the surface tension, and $\rho $ is liquid density.
To estimate  values of neutron cross sections on bubbles in liquid helium we use \cite{bl1} $\beta=1.02 \cdot 10^{-1}\; eV\; nm^{-2}$, $\rho=0.145\; g\; cm^{-3}$ and  $\alpha=2.25 \cdot 10^{-3}\; eV\; nm^{-2}$.  In that case a bubble with a radius $R_0=1.75\; nm$ has frequencies of oscillations $\hbar \omega_b = 3.8\cdot 10^{-5}eV$ and $\hbar \omega_s = 5\cdot 10^{-5}eV$. Since these frequencies have the same order of magnitude, the main difference between total cross sections for the breathing and surface modes comes from the $\Phi$-function in Eqs. (\ref{fbr}) and (\ref{fsf}). Therefore,  neutron cross sections have different energy dependence for breathing and surface modes(see  figure (\ref{ener100}); they also have different dependency on the bubble radius (see  figure (\ref{r175})).

Since the loss (or gain) of energy by neutron at each interaction is equal to $\hbar \omega $, and $\hbar \omega \propto R_0^{-3/2}$, this transferred energy could be adjusted by changing of the radius of the bubble. Due to  theoretically understandable nature of neutron-bubble interactions, very cold and ultra cold neutrons could be to be used to study  bubble formations and dynamics in liquids (for example, bubbles in liquid helium). However, the possible and even more important application of this process is the opportunity of the neutron moderation on the bubble gases. Inelastic scattering of neutrons on  bubbles could slow down thermal neutrons to very cold or ultra cold energies by a small number of interactions without significant loss of the neutron intensity. It could increase the the rate of the production of very cold and ultra cold neutrons by orders of magnitudes. Moreover, using electron bubbles in liquid helium (see, for  example papers \cite{eb1,eb2} and references therein) one can manipulate them \cite{gg} by applying the proper electric and magnetic fields, which would lead to the opportunity \cite{gg} to modulate fluxes  of very cold and ultra cold neutrons.
 At the current stage, however, it is impossible to say how feasible these applications could be. They required   more experimental studies, which are under consideration \cite{vnpr}.

\begin{acknowledgments}
I thank V. V. Nesvizhevsky, who brought to my attention the problem of interactions of UCN with nano-size particles, and J. R. Calarco, G. L. Greene, R. Golub, E. Korobkina, and R. Prozorov for helpful discussions.
This work is supported  by the US Department of Energy, Grant No. DE-FG02-03ER46043.
\end{acknowledgments}


\end{document}